\documentclass[12pt,preprint]{aastex}

\newcommand{\HII}{\mbox{H\thinspace{\sc ii}}} %
\newcommand{\Lsun}{\mbox{$L_{\sun}$}} %

\begin{document}
\title{654~GHz Continuum and $\mathbf{C^{18}O \; (6-5)}$ Observations of G240.31+0.07 with the Submillimeter Array}
\author{Huei-Ru~Chen\altaffilmark{1,2}, Yu-Nung~Su\altaffilmark{2}, Sheng-Yuan~Liu\altaffilmark{2}, Todd~R.~Hunter\altaffilmark{3}, David~J.~Wilner\altaffilmark{3}, Qizhou~Zhang\altaffilmark{3}, Jeremy~Lim\altaffilmark{2}, Paul~T.~P.~Ho\altaffilmark{2,3}, Nagayoshi~Ohashi\altaffilmark{2}, \& Naomi~Hirano\altaffilmark{2}}
\altaffiltext{1}{Institute of Astronomy and Department of Physics, National Tsing Hua University, Hsinchu 30013, Taiwan; hchen@phys.nthu.edu.tw}
\altaffiltext{2}{Institute of Astronomy and Astrophysics, Academia Sinica, P.O. Box 23141, Taipei 10617, Taiwan}
\altaffiltext{3}{Harvard-Smithsonian Center for Astrophysics, 60 Garden Street, Cambridge, MA 02138}

\begin{abstract} 
We report a dual-band observation at 223 and 654~GHz (460~\micron) toward an ultracompact (UC) \HII\ region, G240.31+0.07, with the Submillimeter Array.
With a beam size of $1\farcs5 \times 0\farcs8$, the dust continuum emission is resolved into two clumps, with clump~A coincident well with an $\mathrm{H_2O}$ maser and the UC~\HII\ region. 
The newly discovered clump, B, about 1\farcs3 ($\simeq 8.3 \times 10^3 \; \mathrm{AU}$) to the southwest of clump~A, is also associated with $\mathrm{H_2O}$ masers and may be a more recent star-forming site.
The continuum flux densities imply an opacity spectral index of $\beta = 1.5 \pm 0.3$, suggestive of a value lower than the canonical $2.0$ found in the interstellar medium and in cold, massive cores.
The presence of hot ($\simeq 100 \; \mathrm{K}$) molecular gas is derived by the brightness ratio of  two $\mathrm{H_2CO}$ lines in the 223~GHz band.
A radial velocity difference of $2.5 \pm 0.4 \; \mathrm{km \, s^{-1}}$ is found between the two clumps in $\mathrm{C^{18}O \; (6-5)}$ emission.
The total (nebular and stellar) mass of roughly $58 \; M_{\sun}$ in the central region is close to, but not by far larger than, the minimum mass required for the two clumps to be gravitationally bound for binary rotation.
Our continuum data do not suggest a large amount of matter associated with the $\mathrm{H_2}$ knots that were previously proposed to arise from a massive disk or envelope.
\end{abstract}

\section{INTRODUCTION}
\begin{figure}
\plotone{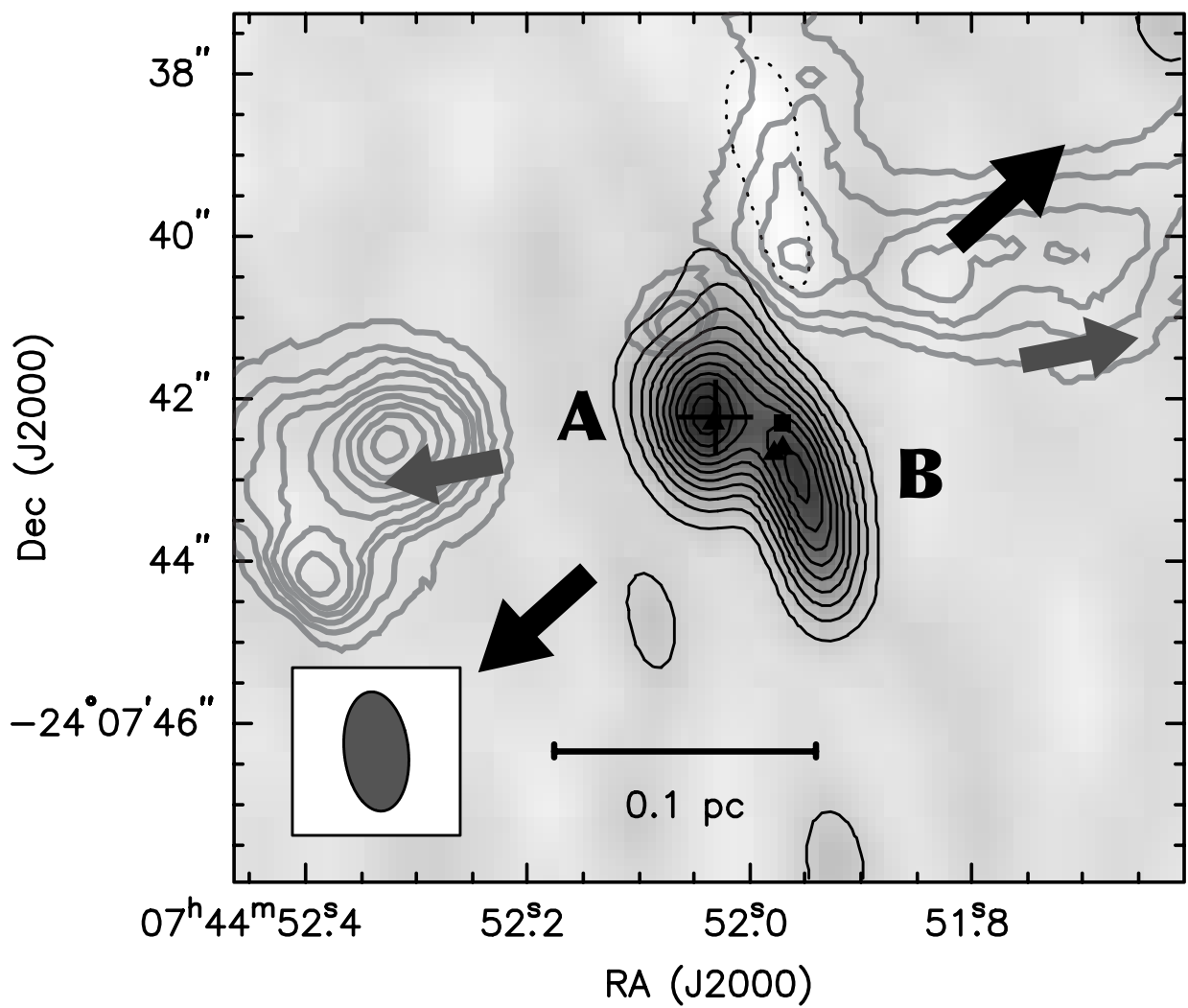}
\caption{SMA 654~GHz line-free continuum map ({\it thin lines}) toward G240 with the $\mathrm{H_2}$ emission features ({\it gray thick lines}; {Kumar} {et~al.} 2003).
The continuum emission is resolved into two clumps, with clump~A coincident well with a VLA 6~cm point source ({\it cross}; {Hughes} \& {MacLeod} 1993).
Triangles represent the locations of $\mathrm{H_2O}$ masers ($0\farcs1$ positional error; {Migenes} {et~al.} 1999) while the square indicates an OH maser spot ($0\farcs4$ postional error; {Caswell} 2003).
Two pairs of arrows indicate the two outflow axes that have been observed in $\mathrm{CO \; (3-2)}$ ({Hunter} 1997): a prominent outflow axis at $\mathrm{P.A. = 138\arcdeg}$ ({\it black arrows}) and a possible second outflow axis at $\mathrm{P.A. = 101\arcdeg}$ ({\it gray arrows}).
Contour levels correspond to ($-2$, $2$, $4$ to $20$)$\times 0.13 \; (1\sigma) \; \mathrm{Jy \, beam^{-1}}$ with a beam size of $1\farcs5 \times 0\farcs8$ ($\mathrm{P.A. = 6\arcdeg}$).
\label{654cont}}
\end{figure}

The bright far-infrared source, G240.31+0.07 (hereafter G240), at a distance of $6.4 \; \mathrm{kpc}$ ({MacLeod} {et~al.} 1998), is known to harbor an ultracompact (UC) \HII\ region associated with $\mathrm{H_2O}$ and OH masers ({Migenes} {et~al.} 1999; {Caswell} 2003).
The flux density at 100~\micron\ suggests a luminosity of $L = 10^{4.7} \, \Lsun$, consistent with a spectral type O8.5 zero-age main sequence (ZAMS) star ({MacLeod} {et~al.} 1998).
Observations with the Very Large Array (VLA) detected a point source with radio free-free emission corresponding to ionization by a ZAMS B0.4 star ({MacLeod} {et~al.} 1998).
On the large scale, the CO~(1-0) emission exhibits broad line wings up to $38.6 \; \mathrm{km \, s^{-1}}$, possibly tracing a bipolar outflow ({Shepherd} \& {Churchwell} 1996).
With improved angular resolution of 20\arcsec, {Hunter} (1997) resolved the CO~(3-2) emission and found a prominent bipolar outflow at a position angle (P.A.) of 138\arcdeg\ and a possible second outflow axis at $\mathrm{P.A.} \simeq 101\arcdeg$.
Each of the two bipolar outflow extends roughly 80\arcsec\ in length.
Subsequent infrared study found two bright, elongated H$_2$ emission features within 16\arcsec\  (0.5~pc) of the UC~\HII\ region (Fig.~\ref{654cont}; {Kumar} {et~al.} 2002). 
{Kumar} {et~al.} (2003) analyzed emission lines of $\mathrm{H_2}$ and [FeII] and concluded the H$_2$ emission being a result of non-dissociative $J$-shocks.
The $\mathrm{C^{18}O \; (2-1)}$ emission, from $65$ to $70 \; \mathrm{km \, s^{-1}}$,  shows a linear velocity gradient of $0.1 \; \mathrm{km \, s^{-1} \, arcsec^{-1}}$ along the H$_2$ axis with a 33\arcsec\ beam.
The authors further argued that the H$_2$ emission features arise in a rotating disk or envelope around G240.

Using single-dish telescopes, millimeter and submillimeter observations toward G240 can not resolve structures smaller than 12\arcsec.
Interferometric observations with high angular resolutions are crucial to investigating the linear velocity gradient seen in the $\mathrm{C^{18}O \; (2-1)}$ emission.
Our current study is enabled by the Submillimeter Array\footnotemark[3] (SMA) ({Ho}, {Moran}, \& {Lo} 2004) with its highest frequency band at 654~GHz.
Dust continuum emission and the higher transition of $\mathrm{C^{18}O \; (6-5)}$ ($E_{up} = 110.6 \; \mathrm{K}$) should better trace warm components in this region.
Moreover, the temperature and the dust properties can be studied by combining the simultaneous observation in the 223~GHz band.

\footnotetext[3]{The Submillimeter Array is a joint project between the Smithsonian Astrophysical Observatory and the Academia Sinica Institute of Astronomy and Astrophysics, and is funded by the Smithsonian Institution and the Academia Sinica.}

\section{OBSERVATIONS AND DATA REDUCTION}
The observation was carried out on Feburary 16, 2005, using 6 elements of the SMA in the dual-band mode, which allows simultaneous operation of the 223 and 654~GHz receivers.
The phase center was toward ($7^{\rm h}44^{\rm m}52\fs0,-24\arcdeg7\arcmin42\farcs6$)(J2000). 
The observing cycle comprised scans of 0736+017, Titan, VY~CMa, and G240, and was repeated every 25 minutes.
The total on-source integration time was about 80 minutes for G240.
Data inspection, bandpass and flux calibrations were done within the IDL superset MIR.
The Jovian moon, Callisto, was used to calibrate the passband as well as the flux density scale.
The flux measurements should be accurate to within 30\% and 15\% at 654 and 223~GHz, respectively.
Details of the observations are listed in Table~\ref{t1}.

A major difficulty for interferometric imaging at 654~GHz is the lack of sufficiently strong quasars that can be used to derive complex gains.
Our strategy is to utilize the nearby (5\arcdeg\ away from G240) $\mathrm{H_2O}$ masers at $658.00655 \: \mathrm{GHz}$ ({Menten} \& {Young} 1995) in VY~CMa to trace atmospheric phase fluctuations.
The $\mathrm{H_2O}$ masers were observed in the same sideband with the $\mathrm{C^{18}O \, (6-5)}$ transition.
Calibration of the 654~GHz band was done in two steps.
First, phase-only gain solutions were derived from the $\mathrm{H_2O}$ masers, whose flux density, however, may change on the timescale of a few hours due to its possibly polarized and highly variable nature.
Without a stable flux density, the masers can not be used for gain amplitude calibration.
Instead, we examined the continuum emission of VY~CMa and found it to be a structureless source of 7~Jy with a $1\farcs5 \times 0\farcs8$ beam.
The peak position of the continuum emission agrees with the maser spot to within 0\farcs06, significantly smaller than the beam. 
We therefore derived complex gains from the continuum emission of VY~CMa to correct temporal amplitude variation.
Gain solutions from the two steps were then multiplied together and applied to the other sources.
Judging from the peak poisitons of Titan (40\arcdeg\ away) and 0736+017 (20\arcdeg\ away), we estimated the astrometry should be good to within 0\farcs3.
Both sidebands, each of a 2~GHz bandwidth, were used to generate a line-free continuum map in a multi-frequency synthesis, resulting a frequency equal to the average of the upper and lower sideband frequencies.
Imaging was performed with the MIRIAD package.

Calibration of the 223~GHz band followed the conventional process by solving complex gains against the quasar 0736+017, whose flux density was assumed to be 1.9~Jy.
The gain solutions were then transferred to the other sources.
A line-free continuum map and channel maps of G240 were generated with uniform weighting.

\section{RESULTS AND DISCUSSION}
\subsection{Morphology and Kinematics \label{sec_c18o}}
The 654~GHz continuum emission of G240 is resolved into two clumps (Fig.~\ref{654cont}), with clump A coincident well with a VLA 6~cm point source ({Hughes} \& {MacLeod} 1993) and an $\mathrm{H_2O}$ maser.
The newly discovered clump, B, is also associated with $\mathrm{H_2O}$ masers and is probably a more recent star-forming site ({Codella} {et~al.} 2004).
The two sources may be responsible for the two bipolar outflows in $\mathrm{CO \; (3-2)}$ emission ({Hunter} 1997). 
The peak of clump~A is at ($7^{\rm h}44^{\rm m}52\fs04$,$-24\arcdeg07\arcmin42\farcs20$)(J2000) with $2.76 \; \mathrm{Jy \, beam^{-1}}$ while that of clump~B is at ($7^{\rm h}44^{\rm m}51\fs96$,$-24\arcdeg07\arcmin42\farcs92$)(J2000) with $2.44 \; \mathrm{Jy \, beam^{-1}}$. 
The projected angular separation between the two peaks is $1\farcs3$, equivalent to $8.3 \times 10^3 \; \mathrm{AU}$, with a P.A. of 57\arcdeg.
No other continuum source is detected above the $3\sigma$ sensitivity of $0.39 \; \mathrm{Jy \, beam^{-1}}$ in the field of view.

\begin{figure}
\epsscale{.80}
\plotone{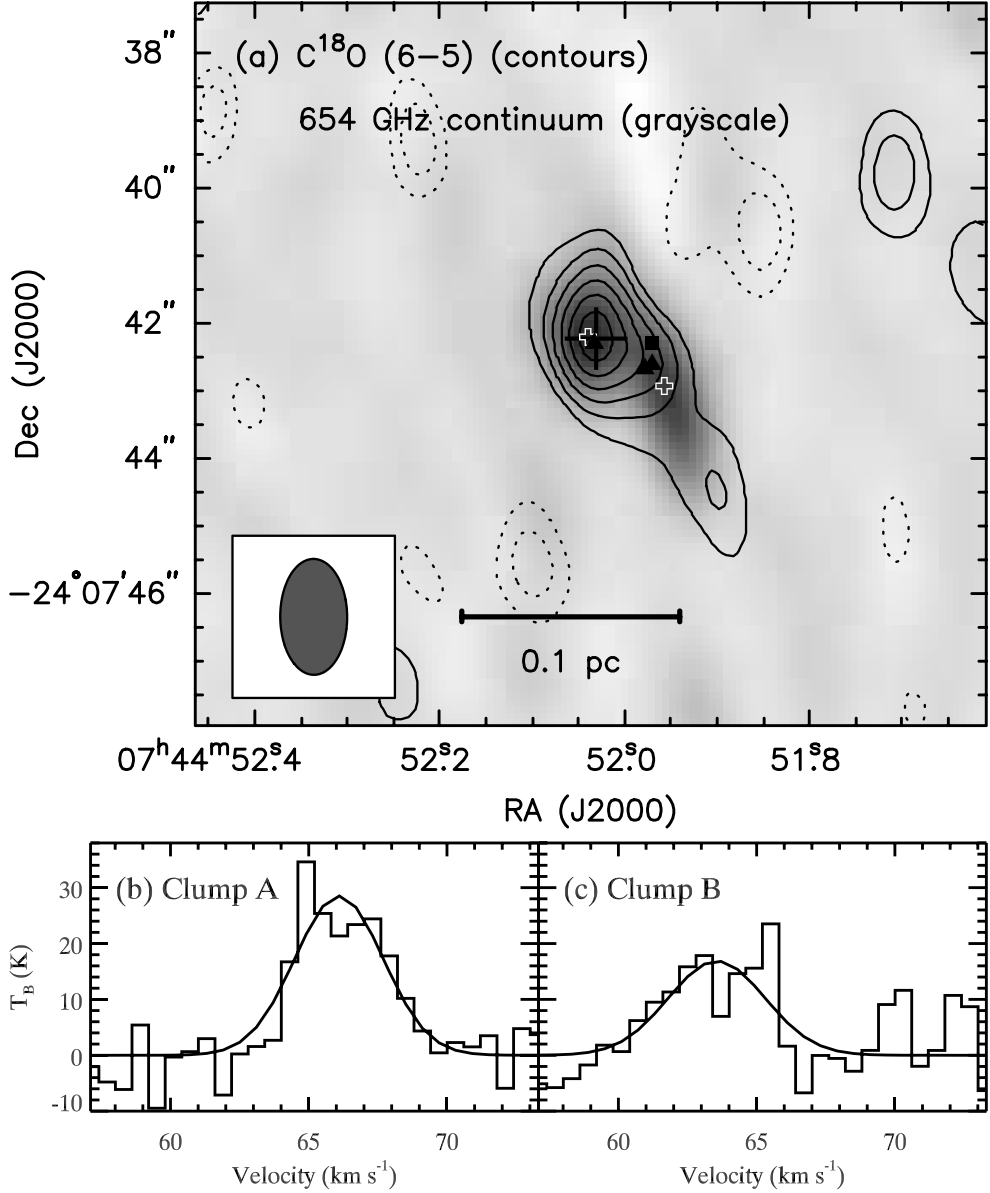}
\caption{({\it a}) $\mathrm{C^{18}O \; (6-5)}$ velocity-integrated brightness map ({\it contours}) from $59.4$ to $69.4 \; \mathrm{km \, s^{-1}}$ superposed on the 654~GHz continuum map ({\it gray scale}).
The $\mathrm{C^{18}O}$ emission are detected toward both continuum clumps.
Symbols follow the use of Fig.~\ref{654cont} with additional white crosses indicating the continuum peaks at 654~GHz. 
Contour levels correspond to ($-3$, $-2$, $2$, $3$ to $7$)$\times 20 \; (1\sigma) \; \mathrm{K \, km \, s^{-1}}$.
({\it b}) and ({\it c}) $\mathrm{C^{18}O \; (6-5)}$ spectra ({\it histograms}) toward clump~A and B with Gaussian fits ({\it solid curves}).
\label{c18o}}
\end{figure}

Fig.~\ref{c18o} shows the $\mathrm{C^{18}O \: (6-5)}$ integrated intensity map and
the spectra toward clump~A and B with Gaussian fits.
Both clumps are associated with $\mathrm{C^{18}O}$ emission and show large Gaussian linewidths attributed to nonthermal effects, e.g. turbulent motions.
The spectrum toward each clump has a Gaussian linewidth of roughly $1.7 \; \mathrm{km \, s^{-1}}$, much larger than the thermal linewidth of $0.17 \; \mathrm{km \, s^{-1}}$ for a clump at $100 \; \mathrm{K}$.
Together with the geometric mean of 0\farcs13 for the emission size, we estimate a virial mass of $68 \; M_{\odot}$ from the $\mathrm{C^{18}O}$ clump. 
The Gaussian peak velocity of clump~A is $\upsilon_A = 66.1 \pm 0.2 \: \mathrm{km \, s^{-1}}$, similar to the ambient velocity of $67.1 \: \mathrm{km \, s^{-1}}$ in the $\mathrm{CO \: (1-0)}$ emission ({Shepherd} \& {Churchwell} 1996).
Clump~B is at $\upsilon_B = 63.6 \pm 0.4 \: \mathrm{km \, s^{-1}}$ based on our $\mathrm{C^{18}O \: (6-5)}$ spectrum and was not detected in the previous $\mathrm{C^{18}O \: (2-1)}$ observation with a larger 33\arcsec\ beam ({Kumar} {et~al.} 2003).
Furthermore, our observation does not detect a velocity gradient along the east-west direction as what has been seen in the large-scale $\mathrm{C^{18}O \: (2-1)}$ emission.
The kinematics at small scales ($0.03 \; \mathrm{pc}$) do not follow the kinematics at large scales ($1.2 \; \mathrm{pc}$).

\subsection{Line Emission in the 223~GHz band}
\begin{figure}
\plotone{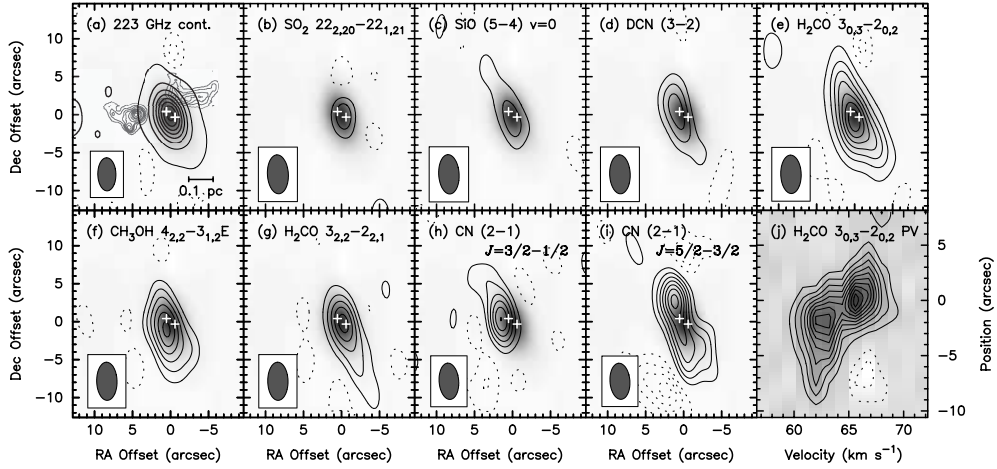}
\caption{({\it a}) 223~GHz line-free continuum map.  Contour levels correspond to ($-4$, $-2$, $-2$, $20$, $40$ to $120$)$\times 2 \, (1\sigma) \; \mathrm{mJy \, beam^{-1}}$ with a beam size of $\mathrm{4\farcs4 \times 2\farcs4}$ ($\mathrm{P.A. = 1\arcdeg}$).   
White crosses indicate the positions of the two 654~GHz continuum peaks. 
The $\mathrm{H_2}$ emission features are shown in gray contours ({Kumar} {et~al.} 2003).
({\it b}) through ({\it i}) Velocity-integrated, from $59.4$ to $69.4 \; \mathrm{km \, s^{-1}}$, brightness maps of lines in the 223~GHz band ({\it contours}) superposed on the 223~GHz continuum map ({\it gray scale}).
Contour levels correspond to ($-8$, $-6$ to $-2$, $2$, $4$ to $14$)$\times 0.8 \, (1\sigma) \: \mathrm{K \, km \, s^{-1}}$ except those in ({\it e}), whose contour levels correspond to ($-4$, $-2$, $2$, $6$ to $26$)$\times 0.8 \; \mathrm{K \, km \, s^{-1}}$.
({\it j}) Position-velocity diagram of a slice through the two continuum peaks, where position $0$ is the center of clump~A.  
Contour levels correspond to ($-4$, $-2$, $2$ to $22$)$\times 0.2 \, (1\sigma) \; \mathrm{K}$.
\label{fig_223lines}}
\end{figure}

Formaldehyde ($\mathrm{H_2CO}$) possesses many transitions of greatly different excitation energies and is a good probe for kinetic temperature ({Mangum} \& {Wootten} 1993).
Two $\mathrm{H_2CO}$ transitions, $3_{0,3}-2_{0,2}$ and $3_{2,2}-2_{2,1}$, with upper state energy levels of $E_{up} = 21$ and $68 \; \mathrm{K}$, respectively, were observed in the 223~GHz band (Fig.~\ref{fig_223lines}{\it e} and \ref{fig_223lines}{\it g}).
We calculate an average temperature in the central $5\arcsec \times 5\arcsec$ region by comparing the mean integrated brightnesses of $13.8 \pm 0.8$ and $4.7 \pm 0.8 \; \mathrm{K \: km \, s^{-1}}$ in the $3_{0,3}-2_{0,2}$ and $3_{2,2}-2_{2,1}$ lines, respectively.
Assuming local thermodynamic equilibrium (LTE), the gas rotational temperature is about $T_g = 96 \pm 35 \: \mathrm{K}$ if both lines are optically thin.

In addition to the two $\mathrm{H_2CO}$ lines, other detected molecular lines are listed in Table~\ref{table_223lines} and velocity-integrated brightness maps are shown in Fig.~\ref{fig_223lines}.
Assuming LTE and optically thin emission, we calculate the abundance for each species (Table~\ref{table_223lines}).
Several lines, e.g. the two $\mathrm{CN}$ lines, are clearly affected by the lack of short baselines, suggesting that a significant fraction of the emission is diffuse.
Although the low angular resolution at 223~GHz can not spatially resolve the two clumps, two distinct kinematic clumps are well separated in velocity space in the $\mathrm{CH_3OH}$ and $\mathrm{H_2CO}$ emissions.
Fig.~\ref{fig_223lines}{\it j}  shows the position-velocity diagram of the $\mathrm{H_2CO}$ emission as an example.

The $\mathrm{SiO} \; (5-4) \: v=0$ emission does not appear to be associated with extensive outflow activities (Fig.~\ref{fig_223lines}{\it c}).  
The derived abundance of $2.4 \times 10^{-11}$ is similar to the average ambient abundances of $10^{-10}$ to $10^{-11}$ in clouds with active star formation ({Codella}, {Bachiller}, \&  {Reipurth} 1999).
There is no clear evidence for $\mathrm{SiO}$ abundance enhancement toward G240 due to outflow activities.

\subsection{Core Properties and Mass Estimate}
The 223~GHz data indicate the presence of extended structure (i.e., a halo), as the visibility amplitudes increase with decreasing baseline length.
To estimate the fluxes of the compact and extended components, we fit the visibilities with a ``core-halo" model consisting of a point source and a one-dimensional Gaussian.
This decomposition gives 223~GHz flux densities of $0.24 \pm 0.04$ and $0.37 \pm 0.06 \; \mathrm{Jy}$ for the core and halo, respectively, and a halo size of about $5\arcsec$.
Similar core-halo dust emission structures have been observed toward other massive star-forming regions ({Su} {et~al.} 2004).
The 654~GHz data is not sensitive to this spatially extended halo.

If the dust emission is optically thin, the continuum flux density $F_{\nu} = \kappa_{\nu} \, M_d \, B_{\nu}(T_d)/D^2$, where $\kappa_{\nu}$, $M_d$, $B_{\nu}(T_d)$, and $D$ are the dust opacity at the observing frequency, $\nu$, the dust mass, the Planck function at a dust temperature, $T_d$, and the distance to the source, respectively.
The opacity spectral index $\beta \equiv d \ln \kappa_{\nu}/d \ln \nu$ can be measured by comparing flux densities at two frequencies.
At 654~GHz, the Rayleigh-Jeans approximation produces 15\% error at a temperature of $100 \; \mathrm{K}$ and is not applied to our calculation.
Because of different $uv$ coverage at 223 and 654~GHz, we compare flux densities in maps made with visibilities of projected baselines within $(22-52) \; \mathrm{k\lambda}$, the range that the two frequencies have in common and corresponds to the compact component within the halo.
The flux densities at 223 and 654~GHz are $0.28 \pm 0.04$ and $10.4 \pm 3.1 \; \mathrm{Jy}$, respectively, and imply that $\beta = 1.5 \pm 0.3$, smaller than the canonical value of $2$ in the interstellar medium and in cold, massive cores ({Hill} {et~al.} 2006).
Small values of $\beta$ have been observed toward several massive star-forming regions ({Chen} {et~al.} 2006; {Beuther} {et~al.} 2004) and may be attributed to grain growth or merely a feature of high-density environment ({Chandler} {et~al.} 1995; {Draine} 2006).
On the other hand, a small $\beta$ can also be a result of the opacity at 654~GHz becoming important.
Assuming a dust opacity of $\kappa_{\mathrm{654 \, GHz}} = 10 \, (\lambda_{\mathrm{mm}}/0.25)^{-1.5} = 4 \; \mathrm{cm^2 \, g^{-1}}$ ({Hildebrand} 1983), the peak intensity of $2.76 \; \mathrm{Jy \, beam^{-1}}$ at clump~A corresponds to a dust column density $0.03 \; \mathrm{g \, cm^{-2}}$, translating to an optical depth $\tau_{\mathrm{654 \, GHz}} \simeq 0.13$.
Hence, the observed small $\beta$ is not due to large opacity at 654~GHz.

The dust mass can be calculated with $M_d = F_{\nu} D^2 / \kappa_{\nu} B_{\nu}(T_g) = 0.47 \; M_{\sun}$ by assuming thermal equilibrium between dust and gas and a 223~GHz dust opacity $\kappa_{\mathrm{223 \, GHz}} = 0.8 \; \mathrm{cm^2 \, g^{-1}}$.
Hence, the total nebular mass is about $47 \; M_{\sun}$ if the gas-to-dust ratio is $100$ ({Savage} \& {Mathis} 1979). 
The estimated virial mass of $68 \; M_{\odot}$ (Sec.~\ref{sec_c18o}) is found to be slightly larger than the nebular mass.  
The cloud may be marginally self-gravitationally bounded or under additional pressure exerted by the outer halo.
The free-free emission of a ZAMS B0.4 star accounts for the minimum stellar mass of $11 \; M_\sun$ ({Hanson}, {Howarth}, \& {Conti} 1997). 
Therefore, the enclosed mass in the central region should be about $58 \; M_\sun$.
Given the radial velocity difference $\Delta\upsilon = 2.5 \pm 0.4 \: \mathrm{km \, s^{-1}}$ and the projected separation $r = 8.3 \times 10^3 \; \mathrm{AU}$, the minimum mass required to hold the two clumps in orbit about each other will be $(\Delta\upsilon)^2 \, r / G = 60 \; M_\sun$.
The total (nebular and stellar) mass in the central region is close to, but not by far larger than, the minimum mass required for the two clumps to be in binary rotation.
We can not rule out the possibility of the two clumps being members of a wide binary system.
Studies of the spectral energy distribution suggested about $390 \; M_\sun$ of matter, mostly in the halo, associated with G240 ({Kumar} {et~al.} 2003).
The two clumps in the central region should be bound inside the gravitational potential well generated by the massive halo.

Neither of our continuum maps shows elongated structures corresponding to the $\mathrm{H_2}$ knots that have been proposed to trace a disk or envelope ({Kumar} {et~al.} 2003).
Assuming typical conditions $\beta \approx 1$ and $T \approx 40 \; \mathrm{K}$ for a disk ({Mannings} \& {Sargent} 1997), we use the 223~GHz continuum sensitivity to place an upper limit on the visual extinction, $A_V$, for this hypothesized disk.
The $2\sigma$ level of $4 \; \mathrm{mJy \, beam^{-1}}$ corresponds to $0.8 \; M_\sun$ within a beam.
The lack of continuum emission suggests that the visual extinction toward the $\mathrm{H_2}$ knots  should be no larger than $4.8 \; \mathrm{mag}$.  
In contrast to a massive disk of $A_V = 10.7 \; \mathrm{mag}$, our data do not suggest such a large amount of matter associated with the $\mathrm{H_2}$ knots.


We are grateful to Ken Young, Rob Christensen, Alison Peck and the SMA team for making the 690~GHz campaign possible.
We also thank Ed Sutton for the discussion on $\mathrm{CH_3OH}$ transitions.


\clearpage

\begin{deluxetable}{lcc}
\tablewidth{0pt}
\tablecolumns{3}
\tablecaption{Parameters for the Dual-band Observation \label{t1}}
\tablehead{\colhead{Frequency band (GHz)} & \colhead{$654$} & \colhead{$223$}}
\startdata
Calibrator & VY~CMa & $0736+017$ \\
Primary beam (arcsec) & $19$ & $56$ \\
Frequency coverage (GHz) & $646.8-648.8$ & $216.5-218.5$ \\
                                            & $656.9-658.9$ & $226.5-228.5$ \\
Typical $T_{sys}$ (K) & $2000$ & $200$ \\
$uv$ range (k$\lambda$) & $22-150$ & $8-52$ \\
Largest visible scale (arcsec) & $4$ & $11$ \\
\hline
\sidehead{Continuum}
Weight scheme & Uniform & Uniform \\
Synthesized beam & $1\farcs5 \times 0\farcs8 \, (6\arcdeg)$ & $4\farcs4 \times 2\farcs4 \, (1\arcdeg)$ \\
rms noise ($\mathrm{mJy \, beam^{-1}}$) & $130$ & $2$ \\
\hline
\sidehead{Line}
Weight scheme & Natural & Uniform \\
Synthesized beam & $1\farcs7 \times 1\farcs0 \, (0\arcdeg)$
                               & $5\farcs0 \times 2\farcs7 \, (2\arcdeg)$ \\
rms noise ($\mathrm{K}$) & $5.3$ & $0.2$ \\
Channel width ($\mathrm{km \, s^{-1}}$) & $0.6$ & $1.2$ \\
\enddata
\end{deluxetable}

\clearpage

\begin{deluxetable}{lllccccc}
\tablewidth{0pt}
\tablecolumns{8}
\tablecaption{Detected Molecular Line Emission \label{table_223lines}}
\tablehead{\colhead{Frequency} & \colhead{Species} & \colhead{Transition} & 
\colhead{$E_{up}$} & \colhead{$T_B^{\: peak}$} &
\colhead{$\int T_B \, d\upsilon$\tablenotemark{a}} & \colhead{$N_\mathrm{X}$} & \colhead{$N_\mathrm{X}/N_\mathrm{H_2}$\tablenotemark{b}} \\
\colhead{(GHz)} & \colhead{(X)} & \colhead{} & \colhead{(K)} & \colhead{(K)} & 
\colhead{($\mathrm{K \: km \, s^{-1}}$)} & \colhead{$\mathrm{(cm^{-2}}$)} & \colhead{} }
\startdata
216.643303 & $\mathrm{SO_2}$ & $\mathrm{22_{2,20}-22_{1,21}}$ & 248 & 0.8 & 4.7 & 1.5E+15 & 4.0E$-9$ \\ 
217.104981 & $\mathrm{SiO}$ & $\mathrm{5-4 \; \upsilon=0}$ & 31 & 1.2 & 4.3 & 8.8E+12 & 2.4E$-11$ \\ 
217.238539 & $\mathrm{DCN}$ & $\mathrm{3-2}$ & 21 & 2.0 & 6.0 & 1.2E+13 & 3.2E$-11$ \\
218.222186 & $\mathrm{H_2CO}$ & $\mathrm{3_{0,3}-2_{0,2}}$ & 21 & 4.3 & 23.5\tablenotemark{c} & 7.1E+14 & 2.0E$-9$ \\
218.440050 & $\mathrm{CH_3OH}$ & $\mathrm{4_{2,2}-3_{1,2} \, E}$ & 45 & 3.0 & 10.1 & 4.1E+15 & 1.1E$-8$ \\
218.475637 & $\mathrm{H_2CO}$ & $\mathrm{3_{2,2}-2_{2,1}}$ & 68 & 1.8 & 8.9 & 7.9E+14 & 2.2E$-9$ \\
226.659543 & $\mathrm{CN}$ & $\mathrm{2 \frac{3}{2} \frac{5/1}{2} - 1 \frac{1}{2} \frac{3/1}{2}}$ & 16 & 1.9 & 9.6\tablenotemark{c} & 3.3E+14 & 9.0E$-10$ \\
226.874764 & $\mathrm{CN}$  & $\mathrm{2 \frac{5}{2} \frac{5/7/3}{2} - 1 \frac{3}{2} \frac{3/5/1}{2}}$ & 16 & 2.6 & 12.3\tablenotemark{c} & 1.6E+14 & 4.6E$-10$ \\
658.553275 & $\mathrm{C^{18}O}$ & $\mathrm{6-5}$ & 111 & 35 & 153 & 6.0E+16 & 1.6E$-7$ \\
\enddata
\tablenotetext{a}{Integrated over channels with emission from $59.4$ to $69.4 \; \mathrm{km \, s^{-1}}$.}
\tablenotetext{b}{Assuming $T_g = 96 \; \mathrm{K}$ and a nebular mass of $47 \; M_{\odot}$, equivalent to $N_{\mathrm{H_2}} = 3.7 \times 10^{23} \; \mathrm{cm^{-2}}$.}
\tablenotetext{c}{Possibly affected by the lack of short baselines and $N_{\mathrm{X}}$ should be considered as a lower limit.}
\end{deluxetable}

\clearpage

\clearpage

\clearpage

\end{document}